\documentstyle[preprint,tighten,aps,epsf]{revtex}

\def\Frac#1#2{\frac{\displaystyle{#1}}{\displaystyle{#2}}}
\def\lsim{\raise0.3ex\hbox{$\;<$\kern-0.75em\raise-1.1ex\hbox{$\sim\;$}}}
\def\gsim{\raise0.3ex\hbox{$\;>$\kern-0.75em\raise-1.1ex\hbox{$\sim\;$}}}
\begin{document}
\preprint{FTUV-12-08, IFIC-00-79}
\title{Supersymmetric origin of a low $a_{J/\psi}$ CP asymmetry}
%\title{Low $a_{J/\psi}$ CP asymmetry with Supersymmetric CP violation in the $K$ system}
\author{A. Masiero, M. Piai}
\address{SISSA -- ISAS, Via Beirut 4, I-34013, Trieste, Italy and \\ INFN,
Sezione di Trieste, Trieste, Italy}
\author{O. Vives}
\address{Departament de F\'{\i}sica Te\`orica and IFIC \\
Universitat de Val\`encia-CSIC, E-46100, Burjassot, Spain}
\maketitle 
\begin{abstract}
We show that general Minimal Supersymmetric extensions of the Standard Model 
(MSSM) allow for a CP asymmetry in $B \rightarrow J/\psi K_S$ well 
below the SM expectations with dominant Supersymmetric contributions to
$\varepsilon_K$ and $\varepsilon^{\prime}/\varepsilon$. Indeed, we provide
an explicit example of an MSSM with non--universal soft breaking terms
fully consistent with the low results of this asymmetry recently announced
by Babar and Belle collaborations.
\end{abstract}

The strong implications of the presence of generic Supersymmetric (SUSY)
soft--breaking terms in Flavor Changing Neutral Current (FCNC) and CP
violation phenomena were readily realized in the early 80's with the beginning
of the SUSY phenomenological studies. The need of an
analog of the GIM mechanism in the scalar sector to suppress 
too large SUSY contributions to $K^0$--$\bar{K}^0$ mixing was emphasized 
\cite{hist} and this showed the large potentiality of looking for SUSY 
signals in FCNC and CP violation experiments. Still, the existence of a single 
experimental measure of CP violation in nature, namely 
indirect CP violation in kaon mixing, $\varepsilon_K$, made it impossible to 
distinguish among a pure Cabibbo--Kobayashi--Maskawa (CKM) origin of CP 
violation or a large supersymmetric (or more generally new physics) 
contribution.
The actual possibility to disentangle these two options arises with the 
comparison of different CP violating processes. Specially, the CP 
asymmetries in $B^0$ decays to be measured in the $B$--factories and the 
improvement of electric dipole moment (EDM) constraints can play a very
important role to accomplish this objective.\\

Recently, the arrival of the first results of $B^0$ CP asymmetries from the
$B$ factories has caused a lot of excitement in the high energy physics 
community.
\begin{equation}
\label{Bfactories}
\begin{array}{ll}
& \\
a_{J/\psi} & = \\ 
& \\
\end{array}
\left\{ 
\begin{array}{ll}
0.12\pm 0.37\pm 0.09 & (\text{ Babar \cite{babar}}) \\ 
0.45 
\begin{array}{l}
{\small +0.43+0.07} \\ 
{\small -0.44-0.09}
\end{array}
& \left( \text{ Belle \cite{belle}}\right) \\ 
0.79 
\begin{array}{l}
{\small +0.41} \\ 
{\small -0.44}
\end{array}
& \left( \text{ CDF \cite{CDF}}\right)
\end{array}
\right.  \label{3ajk}
\end{equation} 
As it is clear from~(\ref{Bfactories}), the errors are still too large to 
draw any firm conclusion. In any case, the Babar and Belle results seem to 
indicate a lower value than the Standard Model (SM) expectations corresponding 
to $0.59\leq \sin \left( 2\beta \right) \leq 0.82$. Several works in the 
literature discussed the possible implications of this small CP asymmetry
\cite{BBtheory,KvsB} and pointed out two possibilities. A small asymmetry
can be due to a large new physics contribution in the $B$ system and/or
to a new contribution in the kaon system modifying the usual determination of 
the unitarity triangle.\\

In a recent letter \cite{KvsB}, two of us proved that, in generic MSSM
models, it is more natural to expect large SUSY contributions to FCNC
and CP--violating observables in the kaon rather then in the $B$
system, unless additional flavor structures thoroughly different from
the fermionic couplings are present in the sfermionic sector
\footnote{Still, even in the absence of additional flavor structures,
SUSY loops can modify the SM contributions proportional to CKM
elements.  However, these changes are relatively small in a GUT
defined MSSM once we take into account all the relevant constraints
\cite{CPcons}.} . In this work we bring those considerations to their
consequences for $a_{J/\psi}$: we show that in non--universal MSSM it
is realistic to reproduce the CP violation in the kaon system through
SUSY effects, while being left with a small $a_{J/\psi}$ in the $B$
system. Indeed the role of the CKM phase could be confined to the SM
fit of the charmless semileptonic $B$ decays and
$B^0_d$--$\bar{B^0_d}$ mixing, while dominantly attributing to SUSY
the $K$ CP violation ($\varepsilon_K$ and
$\varepsilon^{\prime}/\varepsilon$). In this case the CKM phase can be
quite small, hence leading to a low $a_{J/\psi}$ CP asymmetry.\\

Although these results are quite general \cite{KvsB,murayama,EDMfree}, to make 
our point more explicit we prefer to discuss a concrete example based  based 
on type--I \cite{typeI} string theory and, just for the sake of clearness, 
we take a real CKM matrix.\\

We use the model defined in \cite{EDMfree} where soft scalar masses for quark 
doublets and the Higgs fields are all universal at $M_{GUT}$, 
\begin{equation}
\label{doublets}
m_{Q_i}^2=m_{3/2}^2(1-\Frac{3}{2} \cos^2 \theta\ (1- 
\Theta_1^2)),
\end{equation}
while the soft scalar masses for quark singlets are non--universal,
\begin{equation}
\label{singlets}
m_{D_i}^2=m_{3/2}^2(1-3\cos^2 \theta\ \Theta^2_i),
\end{equation}
in terms of $\theta$ and $\Theta_i$ which are 
goldstino angles with the constraint, $\sum \Theta_i^2=1$, and 
$m_{3/2}$ the gravitino mass\footnote{The complementary 
situation with non--universal doublets and universal singlets or the complete
non--universal picture are also perfectly possible. In the literature, 
the case where large flavor--independent soft phases may give a dominant contribution to CP violation has been discussed in \cite{kane}}.\\

In this basis of diagonal sfermion masses, the Yukawa matrices are not
diagonal and they can be written as $v_1\, Y_d =
{K^{D_L}}^\dagger\cdot M_d\cdot K^{D_R}$ and $ v_2\, Y_u =
{K^{U_L}}^\dagger\cdot M_u\cdot K^{U_R}$, with ${K^{U_L}}^\dagger
\cdot K^{D_L} = K_{CKM}$. For definiteness we take the Yukawa matrices
to be hermitian such that $K^{D_R}=K^{D_L}$ and
$K^{U_R}=K^{U_L}$. Moreover, as announced in the introduction, we take
$K_{CKM}$ to be completely real. These matrices, $K^{D_L}$ and
$K^{U_L}$, are completely  unknown unitary matrices, nevertheless, as
discussed in
\cite{KvsB,restrict}, given the smallness of CKM mixings, it is
natural to expect that Yukawa matrices are strongly
hierarchical. Then we may take as a typical situation the case where the
mixings in both
$K^{D_L}$ and $K^{U_L}$ are of the same order as the mixings in
$K_{CKM}$. Notice that, in general, these matrices can have a
different structure, however, a Cabibbo--like mixing among the
first two generations is required to reproduce the CKM
matrix, and this is, indeed the key ingredient in our
discussion. This feature will be shared by any other texture and,
as shown in \cite{KvsB} other mixings have a smaller effect. In any case,
given that now $K^{D(U)_L}$ measure
the flavor
misalignment among, $d(u)_L$--$\tilde{Q}_L$ and we have already used
the rephasing invariance of the quarks to make $K_{CKM}$ real, it is
evident that we can expect new observable (unremovable) phases in the
quark--squark mixings \cite{botella}, and in particular in the first
two generation sector, i.e.,
\begin{eqnarray}
K^{D_L}=\ \left(\begin{array}{ccc} 1 - \lambda^2/2 & \lambda \ e^{i \alpha}  
& A\ \rho\ \lambda^3 e^{i \beta} \\ 
- \lambda \ e^{-i \alpha} & 1 - \lambda^2/2 & A\ \lambda^2 \ e^{i \gamma}\\ 
A\ \lambda^3\ ( e^{- i (\alpha +\gamma)} - \rho\ e^{- i \beta})  & 
\ \ \ -A\ \lambda^2 \ e^{- i \gamma}  & 1 
\end{array}\right)
\label{Kdl}
\end{eqnarray}
to ${\cal{O}}(\lambda^4)$ and $A$, $\rho$ the usual parameters in the 
Wolfenstein parameterization (with $\eta=0$) both them ${\cal{O}}(1)$.\\

Following Ref.~\cite{KvsB} it is straightforward to estimate the $RR$ MI
as, 
\begin{eqnarray}
\label{AMI}
(\delta^{d}_{R})_{i j}&\ =\ \Frac{1}{m^{2}_{\tilde{q}}}\ \Big(
(m_{\tilde{D}_{2}}^2 - m_{\tilde{D}_{1}}^2 )\ K^{D_L}_{i 2} {K^{D_L}_{j
2}}^*\  +\ (m_{\tilde{D}_{3}}^2 - m_{\tilde{D}_{1}}^2 )\ K^{D_L}_{i 3} 
{K^{D_L}_{j 3}}^* \Big)
\end{eqnarray}
However, due to the gluino dominance in the squark eigenstates at $M_W$ we 
can say that $ m^{2}_{\tilde{q}}(M_W)\approx 6\ m_{\tilde{g}}^2(M_{GUT})$. 
In this model \cite{EDMfree}, the gluino mass at $M_{GUT}$ is given by 
$m_{\tilde{g}}^2= 3 m_{3/2}^2 \sin^2 \theta$. 
Replacing these values in Eq~(\ref{AMI}), this means, for the kaon system, 
\begin{eqnarray}
\label{dR12}
(\delta^{d}_{R})_{1 2}&\simeq& \Frac{\cos^2 \theta (\Theta_1^2 - 
\Theta_2^2)}{ 6 \sin^2 \theta}K^{D_L}_{1 2} {K^{D_L}_{2 2}}^*\ +
\Frac{\cos^2 \theta (\Theta_1^2 - \Theta_3^2)}{ 6 \sin^2 \theta}K^{D_L}_{1 3} 
{K^{D_L}_{2 3}}^* \nonumber \\ 
&\simeq & \Frac{\cos^2 \theta (\Theta_1^2 - \Theta_2^2)}{ 6 \sin^2 \theta} \ 
\lambda\ e^{i \alpha}
\end{eqnarray}
This value has to be compared with the MI bounds required to saturate 
$\varepsilon_K$ \cite{MI}, that in this case are, 
$\mbox{Im}(\delta^{d}_{R})_{1 2}^{bound} \leq 0.0032$. Taking 
$\theta \simeq 0.7$ as in
\cite{EDMfree} we get,
\begin{eqnarray}
\label{result}
(\delta^{d}_{R})_{1 2}&\simeq& 0.035 (\Theta_1^2 -\Theta_2^2) \sin \alpha
\lsim 0.0032.
\end{eqnarray}
Hence, it is clear that we can easily saturate $\varepsilon_K$ without 
fine--tuning. Similarly, it is already well--known 
\cite{KvsB,EDMfree,restrict,eps'} that these models can analogously saturate 
$\varepsilon^\prime/\varepsilon$.
In summary, we have explicitly shown that a generic MSSM can fully saturate
the observed CP violation in the kaon system.\\  

Now we turn to the CP asymmetries in the $B$ system. Once more, with 
Eq.~(\ref{AMI}) we have,
\begin{eqnarray}
\label{dR13}
(\delta^{d}_{R})_{1 3}&\simeq& \Frac{\cos^2 \theta (\Theta_2^2 - 
\Theta_1^2)}{ 6 \sin^2 \theta}K^{D_L}_{1 2} {K^{D_L}_{3 2}}^*\ +
\Frac{\cos^2 \theta (\Theta_3^2 - \Theta_1^2)}{ 6 \sin^2 \theta}K^{D_L}_{1 3} 
{K^{D_L}_{3 3}}^* \nonumber \\ 
&\simeq & A\ \lambda^3 \ \Frac{\cos^2 \theta}{6 \sin^2 \theta}\left( - 
(\Theta_2^2 - \Theta_1^2)\ e^{i (\alpha + \gamma)} + (\Theta_3^2 - \Theta_1^2)
\ ( e^{- i (\alpha +\gamma)} - 
\rho\ e^{- i \beta})\right) \lsim 10^{-3},
\end{eqnarray}
to be compared with the MI bound $(\delta^{d}_{R})_{1 3}^{bound} \leq
0.098$ required to saturate the $B^0$ mass difference.  Something
similar can be done in this case for the $LR$ sector \cite{KvsB}. This
implies that the SUSY contribution to $a_{J/\psi}$ is necessarily very
small. Hence, due to the absence of any phase in the CKM mixing
matrix the conclusion is that the CP asymmetry in the $B \rightarrow
J/\psi K_S$ decays will be very small in this extreme (non realistic)
situation with real CKM. It is very interesting to check the
consequences of this picture in other observables, as for instance
rare kaon decays \cite{buras-colang,buras}. Clearly this extreme
situation should be modified with the inclusion of an additional phase
in the CKM matrix, in any case, shifting significantly the usual fit
of the unitarity triangle.\\

In conclusion, we showed that general MSSM schemes allow for a
significantly small $a_{J/\psi}$ CP asymmetry (in agreement with the
present B-factories central values), with the observed CP violation in
the kaon system largely due to new SUSY phases.\\

We thank L. Silvestrini for enlightening conversations.
The work of A.M. was partially supported by the European TMR Project
``Physics across the energy frontier" contract N. HPRN-CT-2000-00148; O.V. 
acknowledges financial support from a Marie Curie EC grant 
(HPMF-CT-2000-00457) and partial support from spanish CICYT 
AEN-99/0692.

\end{document}